\documentclass{aa}
\usepackage{graphicx}
\def\kms{\,km\thinspace s$^{-1}$ }

\def\arcsec{\hbox{$^{\prime\prime}$}}

\def\farcm{\hbox{$.\mkern-4mu^\prime$}}
\def\farcs{\hbox{$.\!\!^{\prime\prime}$}}

\begin{document}


\title{Surface brightness fluctuation distances for nearby dwarf elliptical galaxies
\thanks{Based on observations collected at the Nordic Optical Telescope}
}

\author{Helmut Jerjen\,\inst{1}
\and Rami Rekola\,\inst{2}
\and Leo Takalo\,\inst{2}
\and Matthew Coleman\,\inst{1}
\and Mauri Valtonen\,\inst{2}
} 

\offprints{H.~Jerjen, e-mail: jerjen@mso.anu.edu.au}   

\institute{Research School of Astronomy and Astrophysics, 
The Australian National University, Mt Stromlo Observatory, Cotter Road, Weston ACT 2611, Australia
\and 
Tuorla Observatory, University of Turku, V\"ais\"al\"antie 20, FIN-21500 Piikki\"o, Finland
}

\date{Received 12 July 2001; accepted 9 October 2001}

\abstract{
We obtained deep $B$ and $R$-band CCD images for the dwarf elliptical (dE) galaxies 
DDO 44, UGC 4998, KK98 77, DDO 71, DDO 113, and UGC7356 at the Nordic Optical Telescope. 
Employing Fourier analysis technique we measure stellar $R$-band surface brightness fluctuations 
(SBFs) and magnitudes in 29 different fields of the galaxies. Independent tip of the red giant branch 
distances for DDO 44, KK98 77, DDO 71 are used to convert their set of apparent into absolute SBF 
magnitudes. The results are combined with the corresponding local $(B-R)$  
colours and compared with the $(B-R)-\overline{M}_R$ relation for mainly old, metal-poor 
stellar populations as predicted by Worthey's population synthesis models using Padova isochrones. 
While the colour dependency of the theoretical relation is confirmed by the empirical data, we 
find a systematic zero point offset between observations and theory in the sense that models 
are too faint by $0.13(\pm0.02)$\,mag. Based on these findings we establish a new semiempirical 
calibration of the SBF method as distance indicator for dE galaxies with an estimated 
uncertainty of $\approx$10\%. Taking first advantage of the improved calibration, we 
determine SBF distances for the other three early-type dwarfs UGC 4998, DDO 113, and 
UGC7356. Although found in the M81 group region, previous velocity measurements suggested  
UGC 4998 is a background galaxy. This picture is confirmed by our SBF distance of $10.5(\pm0.9)$\,Mpc. 
We can further identify DDO 113 as a faint stellar system at the near side of the Canes 
Venatici I (CVn I) cloud at a distance of $3.1(\pm0.3)$\,Mpc. The second CVn I member in our 
sample, UGC7356, lies at $6.7(\pm0.6)$\,Mpc and spatially close to NGC 4258 (M 106). We  
derive $BR$ surface brightness profiles and colour gradients for all dwarfs and determine 
photometric and S\'ersic parameters. Finally, we discuss two non-stellar objects in DDO 71 
and UGC 7356 which may resemble globular clusters.
\keywords{
galaxies: clusters: individual:  NGC 2403 group, M81 group, CVn I cloud --
galaxies: dwarf --
galaxies: individual: DDO 44, UGC 4998, KK98 77, DDO 71, DDO 113, UGC 7356 --
galaxies: stellar content --
galaxies: structure
}
}

\titlerunning{ SBF Distances for nearby dE galaxies}
\authorrunning{Jerjen et al.}

\maketitle

\section{Introduction}
In recent years, deep galaxy surveys of the vicinity of the Local Group 
(C\^ot\'e et al.~1997; Karachentseva \& Karachentsev 1998; Jerjen et al.~2000) 
revealed a significant number of very low surface brightness galaxies. 
As these most elusive stellar systems in the 
universe have been identified primarily on morphological grounds their distances remain 
generally unknown. However, many of them are expected to be nearby dwarf 
galaxies and as such preferred targets for studies related to galaxy 
formation and evolution or dark matter. 

While the gas-rich dwarfs, the dwarf irregulars (Irrs), can be located 
relatively easily in space from 21\,cm radio observations (Huchtmeier 
et al.~2000), dwarf elliptical galaxies (hereafter dEs, subsuming ``dwarf 
spheroidals'', see Ferguson \& Binggeli 1994) have a low gas content and 
thus remain undetected in \ion{H}{i}. Moreover, their low surface 
brightness makes optical spectroscopy feasible only for the few brightest 
objects (Jerjen et al.~2000, hereafter JFB00). Hence, the only 
way to identify nearby diffuse dEs and to unveil their physical nature is to 
estimate their distances from their stellar contents.

In principal, the distance of a dE can be obtained via the colour-magnitude 
diagram (Armandroff et al.~1999), the tip of the red giant branch (TRGB) 
magnitude (e.g.~Karachentsev et al.~2000, hereafter K00), or the RR Lyrae 
stars (Saha \& Hoessel 1990). But the requirement of resolving the galaxy into 
stars makes these methods costly and time consuming. A more practical and 
similarly accurate distance indicator would be required if distances for a 
larger number of dE candidates shall be measured in an efficient way. Such 
a distance indicator provides a powerful tool to explore the spatial distribution 
of a statistically meaningful sample of nearby dEs out to a distance $D\approx 10$\,Mpc 
and beyond. As dEs are the best tracers of high-density regions (known as
the morphology-density relation, Binggeli et al.~1990) they flag the 
gravitational centres in the Local Volume and thus hold valuable information 
on the substructure of the Supergalactic plane where most of the nearby 
galaxies are concentrated (e.g.~Jerjen et al.~1998, hereafter JFB98; 
Binggeli 2001 and references therein).

In search for an efficient and accurate distance indicator for dEs, Jerjen
and collaborators (JFB98; JFB00) tested the Surface Brightness Fluctuation (SBF) 
method. This method was introduced by Tonry \& Schneider (1988) to measure distances 
to high surface brightness giant ellipticals. It is based on the discrete sampling of 
a galaxy image with the CCD detector and the resulting pixel-to-pixel variance due 
to the light of {\em unresolved} RGB stars. Analysing CCD data obtained at the 2.3m 
SSO telescope, JFB98 and JFB00 successfully measured $R$-band SBF magnitudes 
in dwarf galaxies found in the nearby Sculptor and Cen\,A groups ($2<D<5$\,Mpc). 

While JFB00 showed convincingly that it is technically feasible to quantify 
surface brightness fluctuations in dEs, there is no {\em empirical} calibration 
of the SBF method as distance indicator for this galaxy type available yet 
due to the lack of calibrators. All reported SBF distances had to rely on the 
theoretical relationship between $(B-R)$ colour and absolute fluctuation magnitude 
$\overline{M}_R$ that was calculated from Worthey's (1994) population synthesis 
models and the Padova isochrones (Bertelli et al.~1994). First results found 
good qualitative agreements between SBF distances for dEs in the Cen\,A group 
and the mean group distance. However, the SBF distance for the Sculptor group 
dwarf ESO540-032 turned out to be significantly shorter than the value derived 
from the RGB tip magnitude (Jerjen \& Rejkuba 2000). The existing results thus 
pose the questions about the reliability of the theoretical models to predict 
$\overline{M}_R$, the accuracy of the SBF method for dEs and about the limits 
of the method. The latter issue is related to the fact that ESO540-032 was 
morphologically classified as an intermediate type dwarf with optical 
properties of both dEs and Irrs. The mixed morphology indicates the presence 
of a more complex underlying stellar population (i.e.~recent star formation 
activities and a wider spread in age and metallicity) than the 
predominantly old, metal--poor populations observed in genuine dEs.

To improve our understanding of the surface brightness fluctuations in dwarf 
elliptical galaxies we studied six nearby dEs in the northern hemisphere. 
DDO 44 (Karachentsev et al.~1999, hereafter K99) is a member of the NGC 2403 
group, UGC 4998 (Bremnes et al.~1998) is a dwarf in the background of the M81 
group, KK98 77 and DDO 71 (K00) are true members of the M81 group, and the two 
dwarfs DDO 113 and UGC7356 are found in the direction of the Canes Venatici I 
(CVn I) cloud (Tully \& Fisher 1987; Binggeli et al.~1990; Bremnes et al.~2000). 
Of particular interest for the present study are DDO 44, KK98 77, and DDO 71 
for which independent TRGB distances have been reported (K98; K00). In 
Table~\ref{tbl1} we give a complete list of our galaxy sample including 
galaxy name, associated group, morphological type within the extended Hubble 
classification system (Sandage \& Binggeli 1984), and coordinates.

\begin{figure*}
\centering
\includegraphics[width=5.6cm]{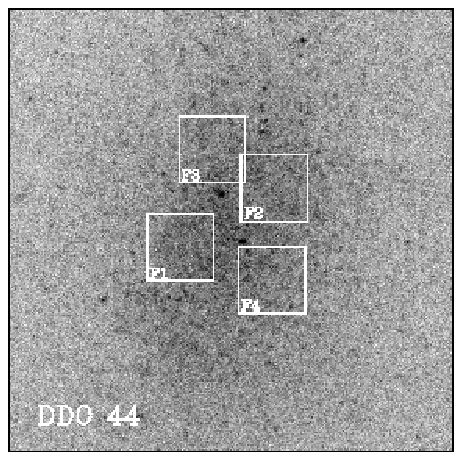}
\includegraphics[width=5.6cm]{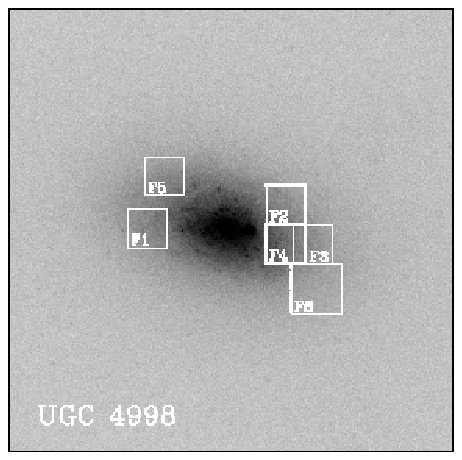}
\includegraphics[width=5.6cm]{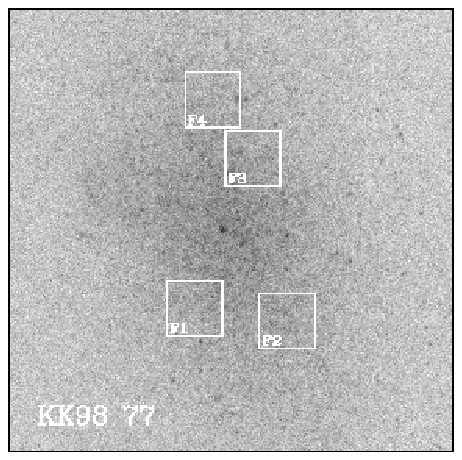}\\
\includegraphics[width=5.6cm]{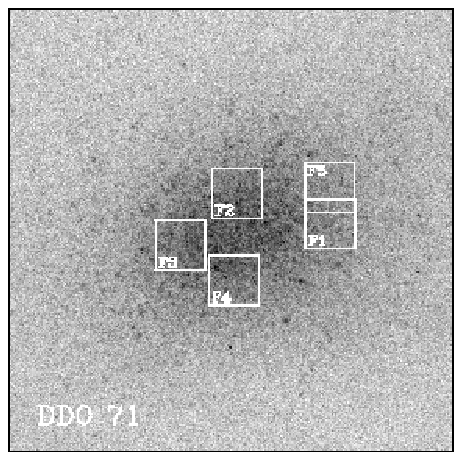}
\includegraphics[width=5.6cm]{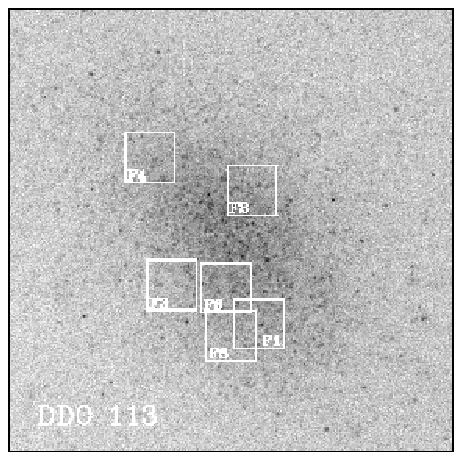}
\includegraphics[width=5.6cm]{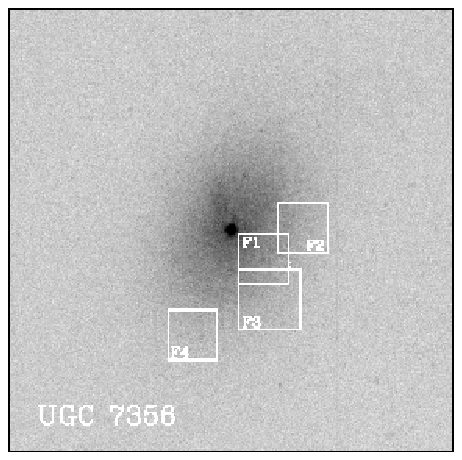}
\caption{Cleaned $R$-band master images of the six dwarf galaxies with the
analysed square SBF fields overlaid. The FOV is $2\farcm5 \times 2\farcm5$. 
North is up, East to the left.} 
\label{fig1}
\end{figure*}

In Sect.~2, we describe the observations and data reduction. The SBF analysis 
is presented in Sect.~3. We develop the semiempirical calibration of the SBF 
method for dEs in Sect.~4 and compare it with the model predictions based on 
synthetic stellar population models. We then discuss the implications 
of our results and derive in Sect.~5 distances for UGC 4998, DDO 113, and 
UGC7356. Finally, we present the integral properties of the dwarfs in Sect.~6  
and draw the conclusions of this work in Sect.~7. 

\begin{table}
\caption{Selected sample of nearby early-type dwarf galaxies\label{tbl1}}
\begin{tabular}{lllcc}
\hline \hline
     &       &      &    R.A.   &    Decl.\\
Name & Group & Type & (J2000.0) & (J2000.0)\\
\hline
DDO44    & NGC2403 & dE   &  07 34 11.4  & 66 53 10  \\
UGC4998  & M81 BG  & dS0  &  09 25 12.1  & 68 22 59  \\
KK98 77 & M81	   & dE   &  09 50 10.5  & 67 30 24  \\
DDO71    & M81	   & dE   &  10 05 06.4  & 66 33 32  \\
DDO113   & CVn I   & dE   &  12 14 57.9  & 36 13 08  \\
UGC7356  & CVn I   & dE,N:&  12 19 09.1  & 47 05 23  \\
\hline \hline
\end{tabular}
\end{table}

\begin{table}
\caption{Summary of observations}
\label{tbl2}
\begin{tabular}{lccccccc}
\hline \hline
      &       &  $t$    &    &      & FWHM \\
Name  &  Date &  (sec)  & F  & $AM$  & ($\arcsec$)\\\hline 
DDO 44  & 22 Jan & 6$\times$600& $B$ & 1.78 & 1.6 \\
DDO 44  & 22 Jan & 7$\times$600& $R$ & 1.53 & 1.3 \\
KK98 77 & 22 Jan & 6$\times$600& $B$ & 1.29 & 1.0 \\
KK98 77 & 22 Jan & 5$\times$600& $R$ & 1.33 & 0.9 \\
DDO 113 & 22 Jan & 6$\times$600& $B$ & 1.01 & 1.1 \\
DDO 113 & 22 Jan & 6$\times$600& $R$ & 1.09 & 0.9 \\
UGC 4998& 23 Jan & 6$\times$600& $B$ & 1.41 & 1.0 \\
UGC 4998& 23 Jan & 6$\times$600& $R$ & 1.32 & 0.9 \\
DDO 71  & 23 Jan & 6$\times$600& $B$ & 1.27 & 0.9 \\
DDO 71  & 23 Jan & 6$\times$600& $R$ & 1.27 & 0.8 \\
UGC 7356& 23 Jan & 6$\times$600& $B$ & 1.08 & 1.2 \\
UGC 7356& 23 Jan & 4$\times$600& $R$ & 1.05 & 1.0 \\
\hline\hline
\end{tabular}
\end{table}

\begin{table}
\caption[]{Photometric calibration coefficients}
\label{tbl3}
\begin{tabular}{ccccr}
\hline \hline
Date & F& ZP & $k$ & $c$ \hspace*{0.8cm}\\
\hline
22 Jan & $B$ & $-25.54\pm0.02$ & $0.22\pm0.01$ & $-0.040\pm0.004$ \\
22 Jan & $R$ & $-25.41\pm0.02$ & $0.09\pm0.02$ & $ 0.031\pm0.006$ \\
23 Jan & $B$ & $-25.54\pm0.02$ & $0.22\pm0.01$ & $-0.037\pm0.004$ \\
23 Jan & $R$ & $-25.42\pm0.02$ & $0.09\pm0.02$ & $ 0.033\pm0.006$ \\
24 Jan & $B$ & $-25.53\pm0.02$ & $0.22\pm0.01$ & $-0.042\pm0.003$ \\
24 Jan & $R$ & $-25.41\pm0.02$ & $0.09\pm0.02$ & $ 0.030\pm0.006$ \\
\hline\hline
\end{tabular}
\end{table}

\section{Observations and Reductions}
CCD images were obtained at the 2.5 metre Nordic Optical Telescope on 
the nights of the 22nd-24th of January, 2001.  The used instrument, the 
Andalucia Faint Object Spectrograph and Camera (ALFOSC), is equipped 
with a 2048 $\times$ 2048 Loral/Lesser CCD detector with a pixel size 
of 15 $\mu$m and a plate scale of 0\farcs188, providing a field of view
6\farcm4 on a side.  The conversion factor is set at 1 $e^- /$ADU. A 
series of four to seven images were taken in the two $B$ and $R$ passbands 
for each of the six dwarf galaxies, along with bias frames, twilight 
flats and photometric standard star fields through the nights.  The 
observing log is given in Table~\ref{tbl2}. The exposure time for an 
individual science frame was 600 seconds. Seeing was ranging 
from 0\farcs7 to 1\farcs6 and all three nights provided photometric conditions.

Image reduction was accomplished using routines within the 
IRAF\footnote{IRAF is distributed by the National Optical Astronomy 
Observatories, which is operated by the Association of Universities 
for Research in Astronomy, Inc., under contract with the National 
Science Foundation} program. We removed the bias level from the
images by using the bias frames and the overscan region of each image.
The images were subsequently trimmed by 50 pixels to remove non-essential 
data from the border.  Finally, each object image was divided by the 
corresponding median combined masterflat.
Photometric calibration was achieved using the Landolt (1992) standard 
star fields regularly observed during each night.  Aperture 
photometry results for each of the standard stars were compared with their 
Landolt magnitudes.  This allowed to determine the photometric zero point 
(ZP), atmospheric extinction coefficient ($k$) and colour term ($c$) 
for each passband and night.  Further analysis revealed stable 
extinction coefficients (variations $< 5\%$) throughout the observation 
period.  The mean $k$ value was calculated for each passband and the 
corresponding values of ZP and $c$ were re-evaluated under this constraint. 
The results are summarised in Table~\ref{tbl3}. The zero points were 
accurate to 0.02\,mag. 

$BR$ images from a galaxy were registered by matching the positions 
of $\approx 50$ reference stars on each CCD frame using {\it starfind}, 
{\it xyxymatch} and {\it imalign}. The sky background level was estimated 
by fitting a plane to selected star-free areas distributed uniformly over 
the CCD area but well away from the galaxy. The sky-subtracted images 
taken in the same passband were cleaned from cosmic rays with {\it crreject} 
and averaged with {\it imcombine} to increase the signal-to-noise. Finally, 
the resulting master images were flux calibrated.

\begin{figure*}[ht]
\centering
\includegraphics[width=17cm]{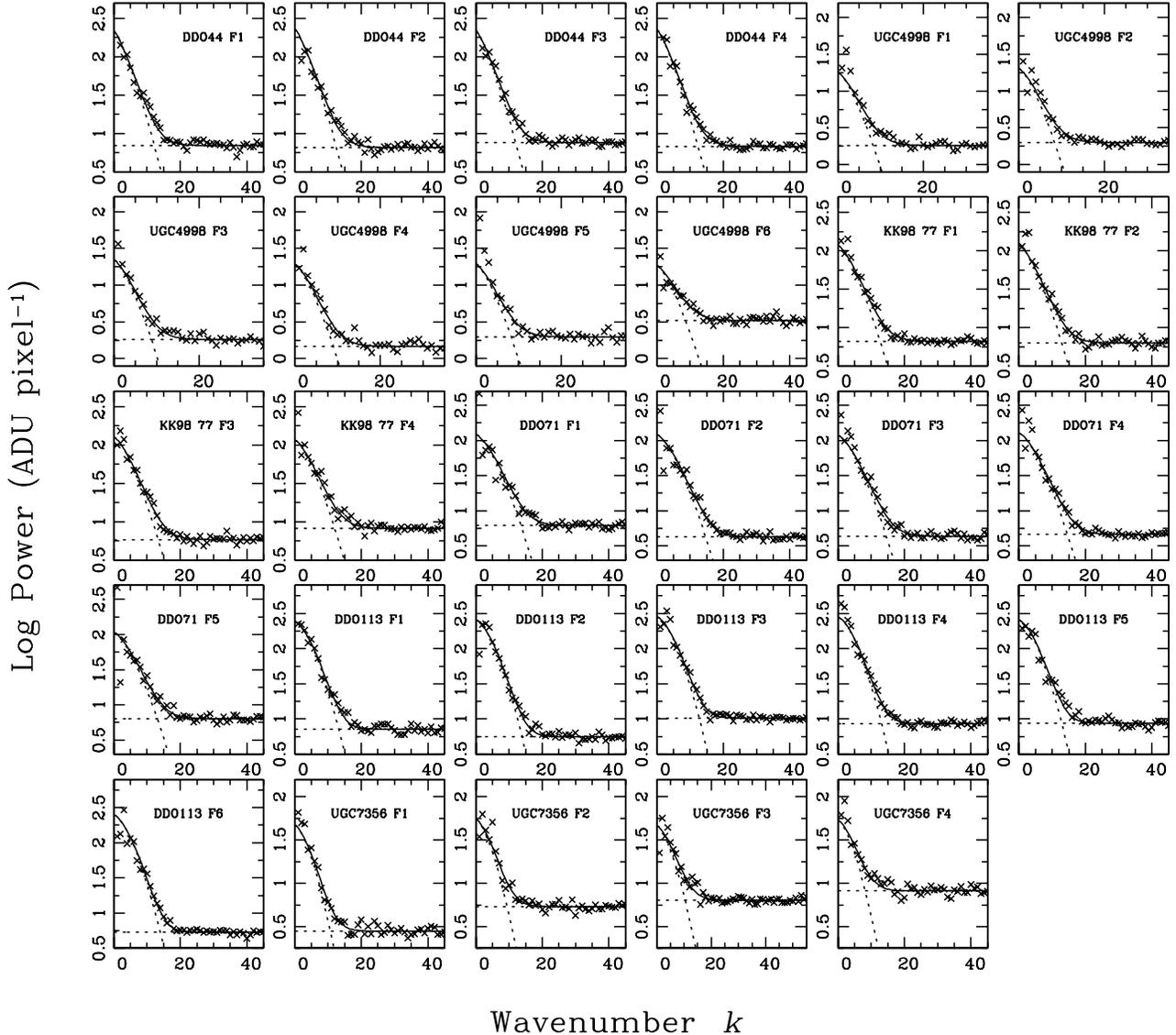}
\caption{
Power spectra (crosses) were measured for 29 galaxy fields. The solid lines in each 
panel represent the least-squares fits to the data. The dashed lines correspond 
to the two components of a fit, a scaled version of the PSF power spectrum 
$P_0\times \mbox{PS}_{\mbox{star}}(k)$ and an additive constant $P_1$. 
}
\label{fig2}
\end{figure*}

\begin{table*}
\caption{Parameters of the SBF analysis \label{tbl4}}
\begin{tabular}{ccccccccc}
\hline\hline
       &  size     & $m_1$  & $\overline{g}$ & $s$ & $P_0$  & $P_1$  & $S/N$ & $P_{\rm BG}/P_0$\\
 Name  & (pixels)  & (mag)  & (ADU)          & (ADU)         & (ADU s$^{-1}$ pixel$^{-1}$) & (ADU s$^{-1}$ pixel$^{-1}$)&           &          \\
 (1)& (2) & (3) & (4)& (5) & (6) & (7) & (8)  & (9) \\
\hline
DDO44 \dotfill F1   & 120 &25.24       &  39.9&$1489.7\pm0.3$ & $0.359\pm0.017$ & 0.012 &  22.8    &  0.01 \\ 
\dotfill F2         & 120 &            &  39.8&               & $0.382\pm0.022$ & 0.011 &  25.5    &  0.01 \\ 
\dotfill F3         & 120 &            &  36.7&               & $0.366\pm0.014$ & 0.013 &  21.7    &  0.01 \\ 
\dotfill F4         & 120 &            &  36.5&               & $0.380\pm0.018$ & 0.011 &  25.4    &  0.01 \\ 
UGC4998 \dotfill F1 & 70  &25.27       & 178.9&$1375.6\pm0.2$ & $0.030\pm0.002$ & 0.003 &   4.5    &  0.10 \\ 
\dotfill F2         & 70  &            & 178.8&               & $0.034\pm0.002$ & 0.003 &   5.1    &  0.09 \\ 
\dotfill F3         & 70  &            & 200.7&               & $0.037\pm0.003$ & 0.003 &   5.7    &  0.08 \\ 
\dotfill F4         & 70  &            & 259.1&               & $0.032\pm0.002$ & 0.003 &   5.0    &  0.09 \\ 
\dotfill F5         & 70  &            & 166.6&               & $0.033\pm0.002$ & 0.003 &   5.0    &  0.09 \\ 
\dotfill F6         & 90  &            & 101.5&               & $0.031\pm0.002$ & 0.005 &   3.6    &  0.09 \\ 
KK98 77 \dotfill F1 & 100 &25.26       &  33.5&$1114.5\pm0.2$ & $0.196\pm0.007$ & 0.011 &  15.0    &  0.01 \\ 
\dotfill F2         & 100 &            &  38.3&               & $0.197\pm0.008$ & 0.010 &  16.3    &  0.01 \\ 
\dotfill F3         & 100 &            &  41.6&               & $0.212\pm0.006$ & 0.010 &  17.3    &  0.01 \\ 
\dotfill F4         & 100 &            &  26.9&               & $0.193\pm0.008$ & 0.014 &  12.0    &  0.01 \\ 
DDO71 \dotfill  F1  & 90  &25.28       &  51.3&$1548.0\pm0.3$ & $0.201\pm0.009$ & 0.010 &  16.6    &  0.01 \\ 
\dotfill F2         & 90  &            &  79.2&               & $0.201\pm0.008$ & 0.007 &  22.1    &  0.01 \\ 
\dotfill F3         & 90  &            &  76.3&               & $0.197\pm0.010$ & 0.007 &  17.6    &  0.02 \\ 
\dotfill F4         & 90  &            &  75.1&               & $0.212\pm0.007$ & 0.008 &  20.7    &  0.01 \\ 
\dotfill F5         & 90  &            &  47.2&               & $0.178\pm0.008$ & 0.011 &  12.0    &  0.02 \\ 
DDO113  \dotfill F1 & 90  &25.28       &  30.6&$1170.1\pm0.2$ & $0.399\pm0.018$ & 0.012 &  24.7    &  0.01 \\ 
\dotfill F2         & 90  &            &  48.1&               & $0.423\pm0.019$ & 0.009 &  31.7    &  0.01 \\ 
\dotfill F3         & 90  &            &  25.4&               & $0.471\pm0.019$ & 0.017 &  21.5    &  0.01 \\ 
\dotfill F4         & 90  &            &  30.7&               & $0.467\pm0.019$ & 0.014 &  24.8    &  0.01 \\ 
\dotfill F5         & 90  &            &  25.3&               & $0.422\pm0.035$ & 0.014 &  22.9    &  0.01 \\ 
\dotfill F6         & 90  &            &  46.3&               & $0.411\pm0.016$ & 0.009 &  31.0    &  0.01 \\ 
UGC7356  \dotfill F1& 90 &25.29        & 117.4&$1420.0\pm0.3$ & $0.079\pm0.005$ & 0.005 &   9.3    &  0.04 \\ 
\dotfill F2         & 90  &            &  63.9&               & $0.091\pm0.005$ & 0.009 &   7.5    &  0.03 \\ 
\dotfill F3         & 110 &            &  54.5&               & $0.077\pm0.003$ & 0.011 &   5.3    &  0.04 \\ 
\dotfill F4         & 90  &            &  38.8&               & $0.089\pm0.005$ & 0.014 &   5.2    &  0.03 \\  
\hline\hline						     
\end{tabular}
\end{table*}

\section{SBF Analysis and Results}
We cleaned the $R$-band master images from foreground stars and background 
galaxies using procedures that follow for most parts the recipes of Tonry \& 
Schneider (1988) and JFB00. Point sources and extended objects brighter 
than the cutoff magnitude $m_c=23.5$\,mag were identified 
using DAOPHOT~II (Stetson 1987). Contaminating objects well away 
from the galaxy were patched with plain sky. If the object 
affected the galaxy light, we replaced its area with an adjacent 
uncontaminated patch from the same surface brightness range. 

For the SBF analysis, we modelled the light distribution of 
the cleaned galaxy image using an isophote fitting routine that allows
the centre, ellipticity, and position angle to vary. The best 2D-model 
of the galaxy was subtracted from the {\em original} master image 
and the residual image divided by the square root of the model for 
noise normalization. A number of square subimages (hereafter SBF fields) 
were then defined on the fluctuation image within the 26.5 $R$\,mag\,arcsec$^{-2}$ 
isophotal limit and from the parts of the galaxy that contained only 
small numbers of previously identified contaminating objects. 
The few affected pixels in a SBF field were manually replaced with randomly 
selected galaxy patches outside of the field and in the same surface 
brightness range. The fraction of pixels patched in this way was always less 
than 5\% of the SBF field area. In total, 29 SBF fields were studied. Their sizes 
and their locations on the galaxy images are shown in Fig.~\ref{fig1}. Assuming an 
average seeing of $\approx 1$\,arcsec, a field carried the SBF signal from $200$ to $600$ 
independent points. The overlap between different fields was kept minimal 
($<5$\%) to get a set of independent SBF measurements for each galaxy. 

The next step in the process was to cut out a cleaned SBF field 
from the fluctuation image and to compute its Fourier transform 
and azimuthally averaged power spectrum (Fig.~\ref{fig2}). From 
isolated bright stars on the master image we determined the point spread 
function (PSF) profile. We then fitted a linear combination of the 
flux normalized and exposure time weighted PSF power spectrum and a constant 
at the observed galaxy power spectrum $\mbox{PS}(k)=P_0\,\mbox{PS}_{\mbox{star}}(k) + P_1$,
demanding a least squares minimization.
Data points at low spatial frequencies ($k<4$) were omitted as they are 
likely to be affected by imperfect galaxy model subtraction. Table~\ref{tbl4}
lists all quantities from the SBF analysis. The SBF 
field numbers together with the galaxy name are given in col.~[1].  
The pixel sizes of the SBF fields are given in col.~[2]. The magnitude 
$m_1$ of a star yielding 1 ADU per second on the CCD is listed in 
col.~[3]. $P_0$ (col.~[6]) and $P_1$ (col.~[7]) are the exposure time normalized 
amplitude of the best least squares fit at wave number $k=0$ and the scale-free 
white noise component in the power spectrum, respectively. In the case of our sky-limited
exposures the latter is determined by the ratio of the sky brightness (col.~[5]) 
and the mean galaxy surface brightness within the SBF field (col.~[4]). 
We estimated the uncertainty of the sky level (col.~[5]) by varying the 
galaxy growth curve (see Sect.~6). The quoted error for $P_0$ in 
col.~[6] is the fitting error. 

To estimate the relative contribution to the measured fluctuation power 
from background galaxies below the cutoff magnitude $m_c=23.5$\,$R$\,mag 
we employed equation (12) from Jensen et al.~(1998). That formula is based 
on the assumption of a power-law distribution for background galaxies 
$n(m)=A\,10^{\gamma(K-19)}$ where $A=10^4$ galaxies deg$^{-1}$ 
mag$^{-1}$ at $K=19$ and $\gamma=0.3$ (Cowie et al.~1994). The original 
equation for the $K$-band was adjusted to work in the $R$-band by assuming 
a typical galaxy colour of $(R-K)=2.25$ (de Jong 1996): 

$$P_{\rm BG}={{p^2} \over {(0.8-\gamma)\ln 10}}10^{0.8(m_1-m_c)-\gamma(29.38+R-K-m_c)}$$

\noindent where $p$ is the plate scale. We calculated $P_{\rm BG}$ and determined
for each SBF field the signal-to-noise S/N$=(P_0-P_{\rm BG})/(P_1+P_{\rm BG})$ (col.~[8]) 
and the ratio $P_{\rm BG}/P_0$ (col.~[9]). We found that the 
contribution from unresolved galaxies that remained in the cleaned SBF fields was minimal 
(a few percent) in all but the six fields of UGC 4998 where the portion reached 
a top of 10 percent. 

Globular cluster (GC) systems of target galaxies are another unwanted source 
of fluctuations that can significantly affect the SBF power measured in giant 
ellipticals (Jensen et al.~1998). However this effect is negligible for faint 
dwarf elliptical galaxies as the typical number of GCs is small in such systems. 
It is know from observations that none of the faintest dE companions to the Galaxy 
has any GCs. The lowest luminosity dwarfs in the Local Group containing 
GCs are Fornax ($n = 5$) and Sagittarius ($n = 4$), with $M_B = -12.6$\,mag 
and $M_B = -12.8$\,mag, respectively (Mateo 1998). These results suggest 
that each of our sample galaxies may have a few GCs at most. All would be 
brighter than our cutoff magnitude and thus identified and removed during 
the cleaning process. Two GC candidates are discussed at the end of Sect.~6. 

The stellar fluctuation magnitudes $\overline{m}_R$ were determined with 
$\overline{m}_R=m_1-2.5\log(P_0-P_{\rm BG})$. Moreover, we used the 
cleaned $BR$ galaxy images to determine the $(B-R)$ colour for each individual 
SBF field. Both quantities were corrected for foreground extinction 
using the IRAS/DIRBE maps of dust IR emission (Schlegel et al.~1998). 
The results are summarized in Table~\ref{tbl5}. 

The overall error of $\overline{m}_R^0$ is dominated by the power spectrum fitting 
error which accounts for 3-8\%. Other sources of minor errors are the PSF 
normalisation ($\sim$2\%), the shape variation of the stellar PSF over the CCD 
area (1--2\%) and the uncertainty in the photometric calibration $0.02$\,mag. 
If we adopt a 16\% error for the foreground extinction (Schlegel et al.~1998), 
the formal internal error for a single $\overline{m}_R^0$ measurement is between 
0.06 and 0.12\,mag (col.~[3]). The error associated with the local colour 
$(B-R)_0$ (col.~[4]) has been obtained through the usual error propagation 
formula from the standard errors estimated from the uncertainties of the 
Galactic extinction, sky level and the photometry.

\begin{table}
\caption{Fluctuation magnitudes and colours \label{tbl5}}
\begin{tabular}{cccc}
\hline\hline
\small
       & A$_R$ & $\overline{m}_R^0$&$(B-R)_0$ \\
 Name  & (mag) & (mag)                                  & (mag)    \\
 (1)& (2) & (3) & (4)\\
\hline
DDO44 \dotfill F1   &0.11$\pm0.02$ &26.25$\pm0.09$& 1.05$\pm0.05$\\ 
\dotfill F2         &           &26.18$\pm0.10$& 1.07$\pm0.05$\\ 
\dotfill F3         &           &26.23$\pm0.09$& 1.01$\pm0.05$\\ 
\dotfill F4         &           &26.19$\pm0.10$& 1.08$\pm0.05$\\ 
UGC4998 \dotfill F1 &0.16$\pm0.03$ &29.03$\pm0.06$& 1.13$\pm0.03$\\ 
\dotfill F2         &           &28.88$\pm0.06$& 1.09$\pm0.03$\\ 
\dotfill F3         &           &28.78$\pm0.08$& 1.07$\pm0.03$\\ 
\dotfill F4         &           &28.94$\pm0.06$& 1.07$\pm0.03$\\ 
\dotfill F5         &           &28.93$\pm0.07$& 1.11$\pm0.03$\\ 
\dotfill F6         &           &29.00$\pm0.07$& 1.17$\pm0.03$\\
KK98 77 \dotfill F1 &0.39$\pm0.06$ &26.66$\pm0.10$& 1.20$\pm0.05$\\ 
\dotfill F2         &           &26.65$\pm0.10$& 1.15$\pm0.05$\\ 
\dotfill F3         &           &26.57$\pm0.09$& 1.12$\pm0.05$\\ 
\dotfill F4         &           &26.67$\pm0.11$& 1.09$\pm0.05$\\ 
DDO71 \dotfill  F1  &0.26$\pm0.04$ &26.78$\pm0.09$& 1.25$\pm0.05$\\ 
\dotfill F2         &           &26.78$\pm0.07$& 1.22$\pm0.05$\\ 
\dotfill F3         &           &26.80$\pm0.08$& 1.26$\pm0.05$\\ 
\dotfill F4         &           &26.72$\pm0.07$& 1.23$\pm0.05$\\ 
\dotfill F5         &           &26.91$\pm0.09$& 1.26$\pm0.05$\\ 
DDO113  \dotfill F1 &0.06$\pm0.01$ &26.23$\pm0.09$& 1.06$\pm0.04$\\ 
\dotfill F2         &           &26.16$\pm0.07$& 1.00$\pm0.04$\\ 
\dotfill F3         &           &26.04$\pm0.10$& 0.95$\pm0.04$\\ 
\dotfill F4         &           &26.05$\pm0.09$& 0.93$\pm0.04$\\ 
\dotfill F5         &           &26.16$\pm0.12$& 1.06$\pm0.04$\\ 
\dotfill F6         &           &26.19$\pm0.07$& 1.00$\pm0.04$\\ 
UGC7356  \dotfill F1&0.05$\pm0.01$ &28.05$\pm0.07$& 1.15$\pm0.03$\\ 
\dotfill F2         &           &27.87$\pm0.07$& 1.15$\pm0.04$\\ 
\dotfill F3         &           &28.07$\pm0.07$& 1.14$\pm0.04$\\ 
\dotfill F4         &           &27.90$\pm0.10$& 1.07$\pm0.04$\\  
\hline\hline
\end{tabular}
\end{table}

\section{On the empirical calibration of the SBF method for dEs}
As we pointed out in the introduction, existing SBF distances for nearby dEs 
(JFB98; JFB00) had to rely on results from synthetic stellar population models as no 
calibrator galaxies with independent accurate distances were available. 
To calibrate the apparent stellar fluctuation magnitudes $\overline{m}_R$, 
JFB00 used $\overline{M}_R$ predictions from Worthey's (1994) 
models\footnote{http://199.120.161.183/$\sim$worthey/dial/dial\_a\_pad.html}
combined with the Padova isochrones (Bertelli et al.~1994). For that
purpose, the theoretical relationship between absolute fluctuation magnitude 
$\overline{M}_R$ and the stellar population's integrated $(B-R)$ colour
was computed for a number of single burst and simple composite stellar populations. 
In the first series, the populations covered the \{age=8, 12, 17 Gyr\} $\times$ 
\{[Fe/H]=$-1.7$, $-1.6$, ..., $-1.0$, $-0.5$, $-0.25$, $0$\} parameter space 
(with [Fe/H]$\geq-1.3$ in the case of 17\,Gyr due to model limitations). In 
the second series, the previously defined populations were mixed at the 10, 
20 and 30\% level in mass with a 5\,Gyr old, solar metallicity population.  
In all cases we assumed a Salpeter IMF. The derived SBF distances for five 
dEs studied in the Cen A group (JFB00) indicated good quantitative agreement 
between models and observations. But for ESO540-032, an intermediate type 
(dE/Irr) dwarf in the Sculptor group, a significant difference between the 
SBF (JFB98) and TRGB distances was found (Jerjen \& Rejkuba 2000). From 
the existing results it remains unclear whether the observed discrepancy 
in the latter case is due to some problems with the models or the SBF method 
is just not suitable to estimate distances for dwarf galaxies with mixed 
morphology. Obviously an empirical calibration of the SBF method would 
shed some light on this issue.

\begin{figure*}
\centering
\includegraphics[width=7.5cm]{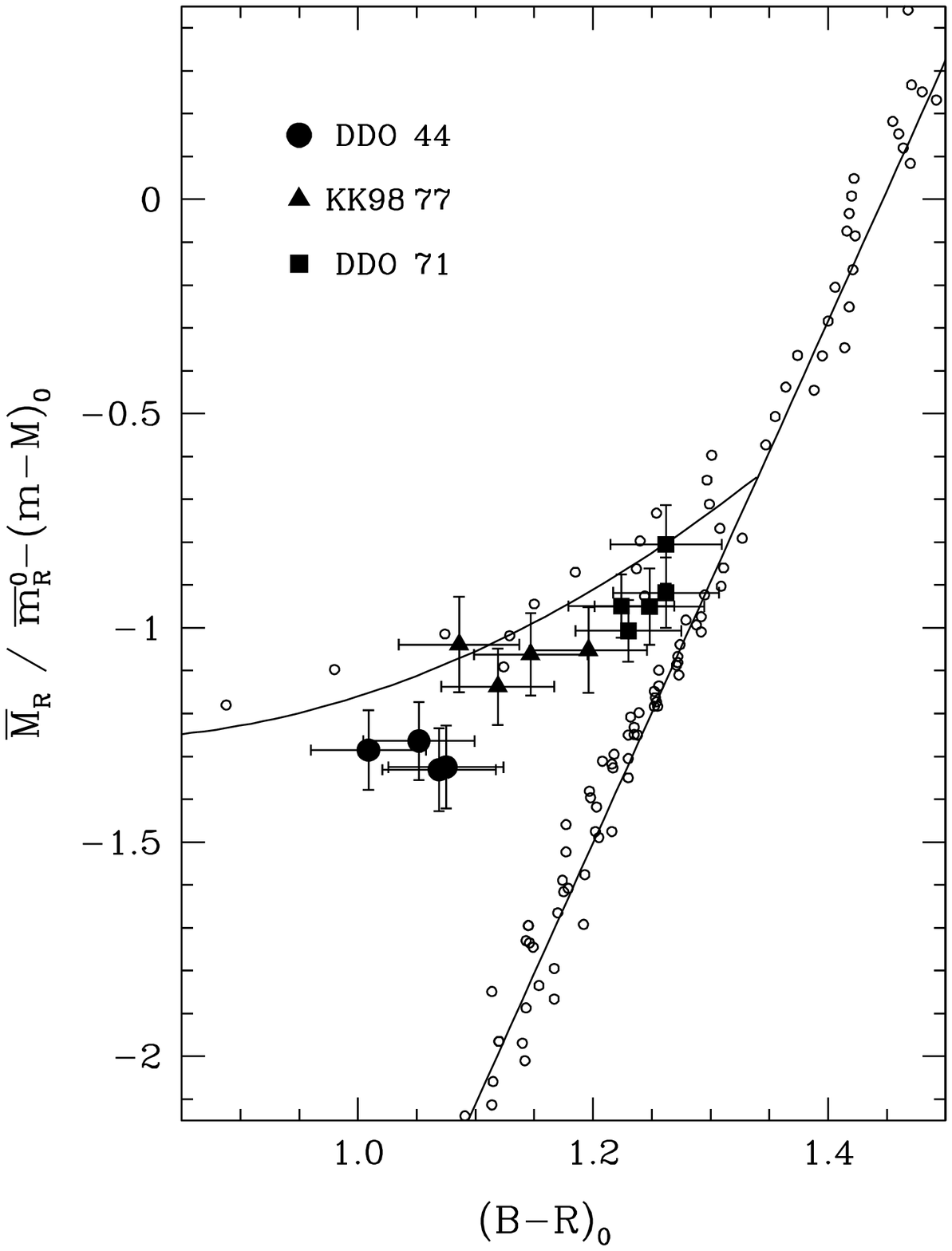}      
\includegraphics[width=7.5cm]{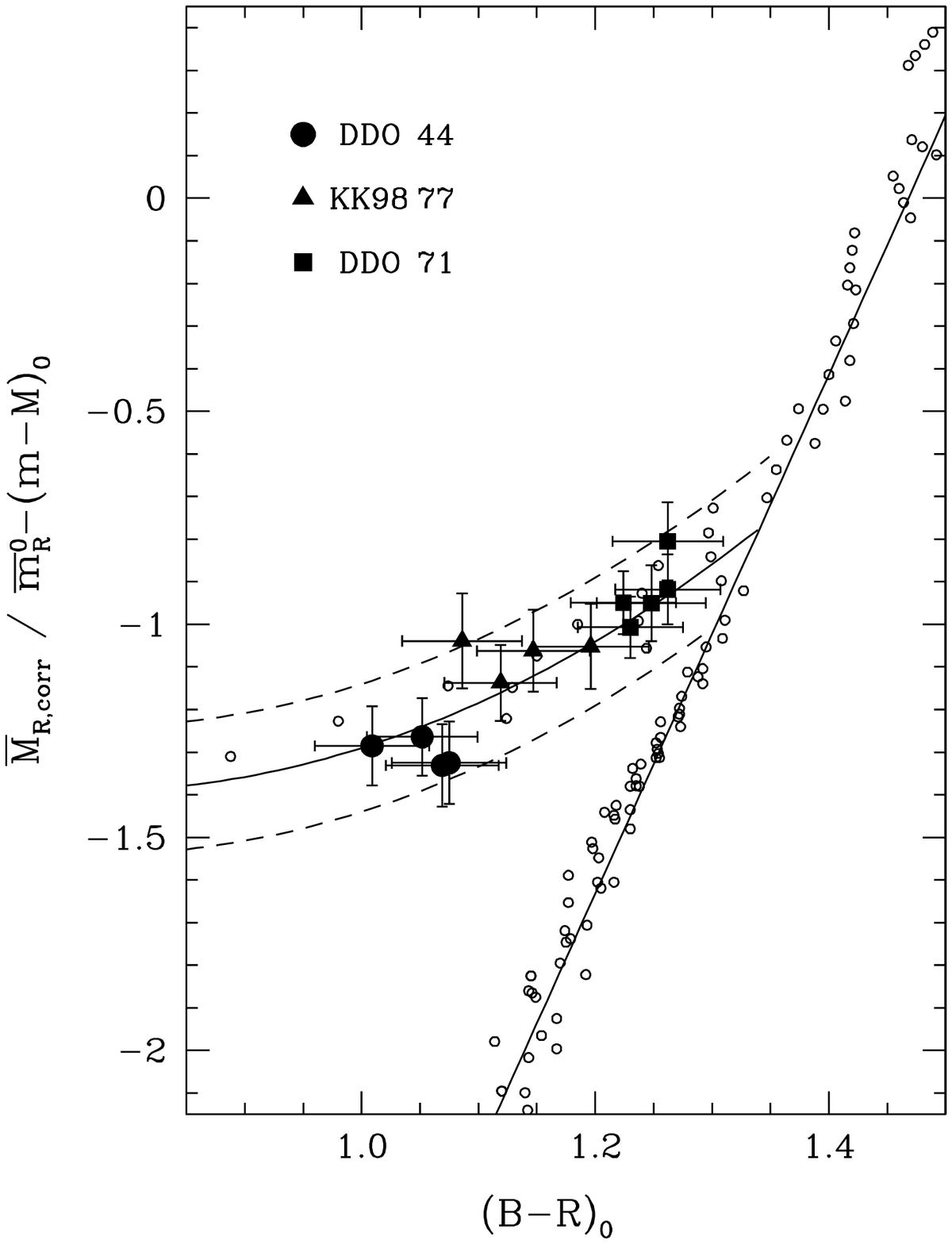}      
\caption{The distribution of absolute SBF magnitudes 
[$\overline{m}_R^0-(m-M)_0$] versus $(B-R)_0$ colours for the 13 SBF fields studied 
in the calibrator galaxies DDO 44, KK98 77, and DDO 71 (filled symbols). 
Superimposed are Worthey+Padova model predictions for a grid of stellar populations 
(open circles): a set of single burst populations that covers the \{age=8, 12, 17 Gyr\} $\times$ 
\{[Fe/H]=$-1.7$, $-1.6$, ..., $-1.0$, $-0.5$, $-0.25$, $0$\} parameter 
space (with [Fe/H]$\geq-1.3$ in the case of 17\,Gyr due to model limitations)
and a set of composite populations where the previously defined populations 
were mixed at the 10, 20 and 30\% level (in mass) with a second generation 
of 5\,Gyr old stars with solar metallicity. The solid lines are the best least 
squares fits to the two branches exhibit by the 116 model data points. A 
colour independent offset of $0.13(\pm0.02)$\,mag is found between theory and 
observations. The two panels show the model data before (left) and after (right) 
the offset correction. The dashed lines above and below the parabolic branch 
indicate the $\pm 0.15$\,mag strip that envelopes the scatter of the data points 
from the different stellar populations and the observed scatter in the fluctuation 
measurements.}
\label{fig3}
\end{figure*}

Only recently Karachentsev and collaborators (K99; K00) published TRGB distances 
for three of our sample galaxies DDO 44, KK98 77, and DDO 71 (Table~\ref{tbl7}). 
The TRGB method has been proven to be a reliable and accurate distance indicator 
for old and metal-poor stellar populations (Da Costa \& Armandroff 1990; 
Lee et al.~1993) such as observed in early-type dwarfs. 
Hence, these data are well suited for a test and empirical calibration of the 
SBF method for dEs. We used the TRGB distances to convert the apparent 
fluctuation magnitudes measured in the 13 SBF fields of the three dwarfs 
into absolute magnitudes. The data are plotted versus their corresponding 
$(B-R)_0$ colours in the left panel of Fig.\ref{fig3}. Superimposed are the 
points (open symbols) of the 116 stellar populations which we described 
in the previous paragraph. The solid lines represent the best analytical fits to 
the two branches exhibit by the model data as derived in JFB00:
$\overline{M}_R = 1.89\,[(B-R)_0-0.77]^2-1.26$ (parabolic branch)
and $\overline{M}_R = 6.09\,(B-R)_0-8.81$ (linear branch).

The calibrator data show a similar colour dependency as the parabolic 
branch, the part of the theoretical locus that is most exclusively defined 
by the very old (17\,Gyr), lowly contaminated and metal-poor ([Fe/H]$<-1.0$) 
populations. However there is a clear systematic shift evident between the 
two data sets. We computed this offset for the individual calibrators as well 
as for the combined data set by fitting the analytic form of the parabolic branch to 
the empirical data keeping the zero point as a free parameter. The 
error-weighted fit results are listed in Table~\ref{tbl6}.

\begin{table}
\caption{$\overline{M}_{R}$ difference between theory and observations 
(parabolic branch).\label{tbl6}}
\begin{tabular}{lc}
\hline \hline\vspace{-3mm}\\
     &$\overline{M}_{R,\mbox{th}}-\overline{M}_{R,\mbox{obs}}$ \\
Name & (mag)     \\
\hline
DDO 44   & 0.19$\pm0.05$ \\
KK98 77  & 0.08$\pm0.05$ \\ 
DDO 71   & 0.10$\pm0.04$\\ 
\hline
All      & 0.13$\pm0.03$ \\
\hline \hline
\end{tabular}
\end{table}

The best match for the combined data occured at a offset of 
$0.13\pm 0.03$\,mag in the sense that model magnitudes are 
too faint by that amount. Correcting the analytic form of 
the parabolic branch accordingly yielded a first semiempirical 
calibration for the SBF method as distance indicator for dEs:

\begin{eqnarray}
&&\quad\overline{M}_R = 1.89[(B-R)_0-0.77]^2-1.39.
\end{eqnarray}

\noindent The right panel of Fig.\ref{fig3} illustrates the 
situation between observations and models after the correction 
was applied to all model data.  

A consistency check confirmed that the Padova isochrones predict 
the same TRGB absolute magnitude of $M_I=-4.05(\pm 0.1)$ (see 
Fig.~3 in Da Costa 1998) as Karachentsev et al.~used in their study
to derive the dE distances. Therefore a difference between models and 
observations at that level can be ruled out as a possible explanation for 
the offset. A detailed analysis of the discrepancy lies outside the scope of this 
paper. It is also interesting to note that all our calibrator galaxies 
happened to lie on the parabolic branch. This circumstance prevented 
an empirical test of the zero point and slope of the second component 
of the theoretical locus, the steeply rising linear branch, with our 
data. Again, we have to leave this issue to a future investigation.

\section{SBF distances for UGC 4998, DDO 113 and UGC 7356}
\begin{figure*}
\centering
\includegraphics[width=18cm]{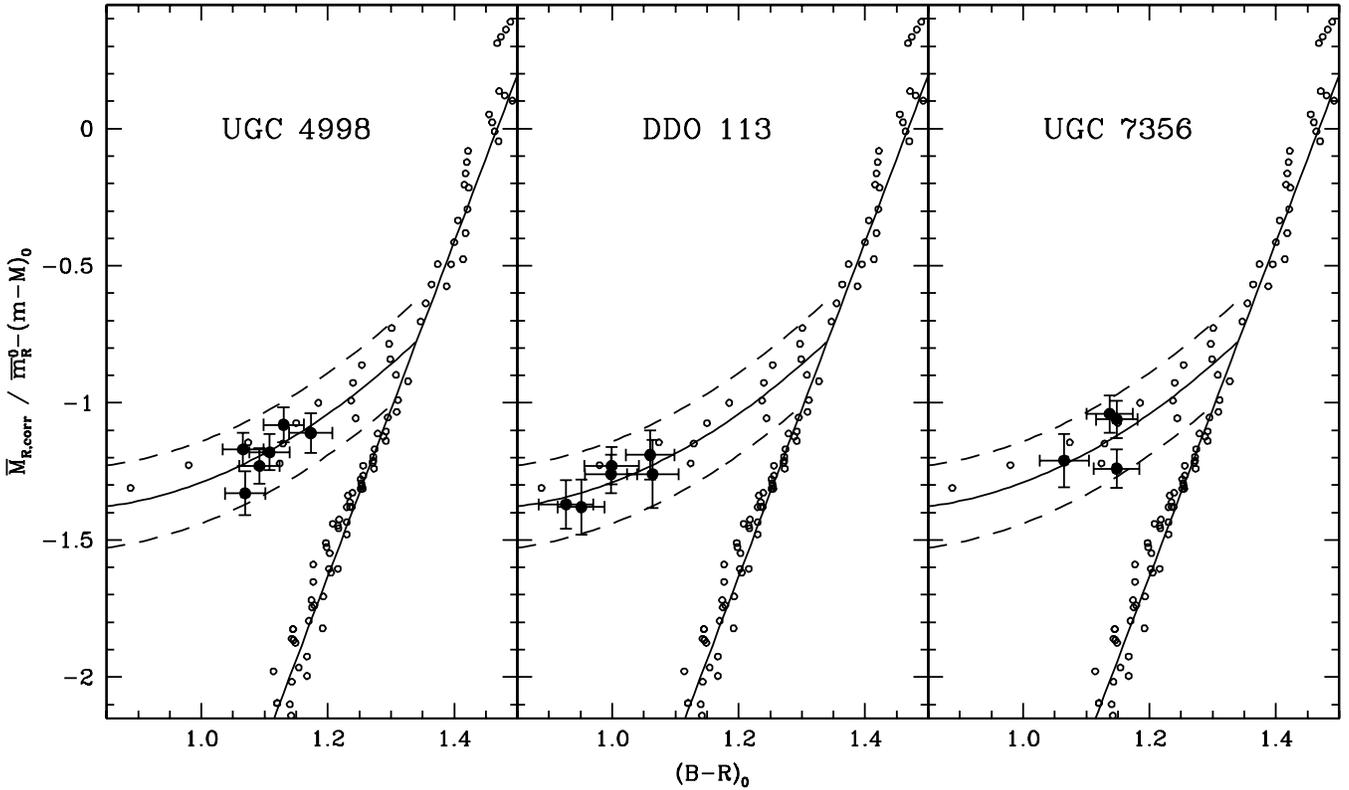}  
\caption{Each set of $[(B-R)_0,\overline{m}_R^0]$ data for a given dwarf
galaxy was fitted to the parabolic branch of the theoretical $(B-R)-\overline{M}_R$ 
relation, using the newly determined empirical zero point. As in the right panel 
of Fig.~\ref{fig3}, the dashed lines below and above the parabolic component of 
the model locus indicate the $\pm 0.15$\,mag uncertainty strip.}
\label{fig4}
\end{figure*}

The improved calibration relation was employed to measure the SBF 
distances of the remaining dwarfs UGC 4998, DDO 113 and UGC 7356. 
All $[(B-R)_0, \overline{m}_R^0]$ data points of a given galaxy were 
shifted simultaneously along the ordinate to get the best least 
squares fit at the parabolic branch of the new calibration (Eq.~1). These fits 
are shown in Fig.\ref{fig4} and the corresponding distance moduli 
for the galaxies listed in Table~\ref{tbl6}. The cumulative error 
from the uncertainty in a single SBF magnitude measurement, the 
calibration function fitting error, and the uncertainty in the calibration 
zero point due to the intrinsic scatter of the model values is 
$\approx 0.15$\,mag. Further adopting an uncertainty of $0.1$\,mag in the 
TRGB zero point (Lee et al.~1993) amounts to the quoted overall uncertainty in 
a SBF distance of $0.18$\,mag.

\begin{table}
\caption{Distances for six nearby early-type dwarf galaxies\label{tbl7}}
\begin{tabular}{lccl}
\hline \hline
     & $(m-M)_0$  &   Distance   & \\
Name & (mag)      & (Mpc) & Reference         \\
 (1) & (2)        & (3)   & (4)       \\
\hline
DDO 44    & 27.52$\pm0.15$ & 3.2$\pm0.2$ & K99           \\
UGC 4998  & 30.11$\pm0.18$ & 10.5$\pm0.9$  & present paper\\
KK98 77   & 27.71$\pm0.15$ & 3.5$\pm0.3$ & K00           \\ 
DDO 71    & 27.72$\pm0.15$ & 3.5$\pm0.3$ & K00           \\ 
DDO 113   & 27.44$\pm0.18$ & 3.1$\pm0.3$ & present paper \\
UGC 7356  & 29.12$\pm0.18$ & 6.7$\pm0.6$ & present paper \\
\hline \hline
\end{tabular}
\end{table}

UGC 4998 is a dwarf S0 galaxy found in the region of the nearby M81 group. 
The mean distance of the group is $\approx 3.7$\,Mpc (K00). Before the 
radial velocity of UGC 4998 was known, this galaxy was believed to be a 
group member. However, the measured velocity of $v_\odot=623$\kms (Falco 
et al.~1999) was significantly higher than the velocity of the group 
centroid $\approx 142$\kms and thus rendered UGC 4998 a background object. 
This picture is now confirmed with the SBF distance of $10.5(\pm0.9)$\,Mpc. 

DDO 113 and UGC 7356 are among the few early-type dwarfs found in the Canes 
Venatici I (CVn I) cloud (Binggeli et al.~1990). This loose galaxy association 
without an obvious concentration covers a huge area in the sky and is known to 
have a considerable depth as estimated from its velocity distribution. Thus it comes 
as no surprise that the two dwarfs span a wide range in distance. While 
DDO 113 is located at the near side of the CVn I cloud with a SBF distance 
of $3.1(\pm0.3)$\,Mpc, UGC 7356 is found at the far side at $6.7(\pm0.6)$\,Mpc
[$(m-M)_0=29.12$]. The latter dwarf is only 5 arcmin away from the southern spiral 
arm of NGC 4258 (M 106), a well-known nearby galaxy with a distance modulus 
measured from observations of circumnuclear masers at $(m-M)=29.29$ 
(Herrnstein et al.~1999) and of Cepheids at $(m-M)=29.54$ (Maoz et al. 1999). 
Our result suggests that UGC 7356 is also spatially close to NGC 4258. Yet the differences in 
velocity (272\kms versus 448\,km\thinspace s$^{-1}$) makes the dwarf an unlikely 
satellite of the giant spiral. In this context another early-type dwarf NGC 4248 
is a more promising candidate for a NGC 4258 companion having a heliocentric velocity 
of 484\kms and a projected distance of 29\,kpc from the spiral centre. A SBF analysis 
of NGC 4248 is planned to confirm the physical relation of the pair.

\section{Integrated properties}
The cleaned, registered, and sky-subtracted $BR$ images of the galaxies (Sect.~3) 
were further analysed to obtained accurate broadband photometry, colour and surface 
brightness profiles. Using the IRAF command {\it ellipse}, the isophotes 
out to the 26.5\,mag arcsec$^{-2}$ limit on the $R$-band image were fitted with 
ellipses to find the radial dependency of the centre coordinates, position angle 
and ellipticity. All parameters turned out to be stable within the errors and a 
set of mean values were calculated for each galaxy. Keeping the four parameters 
constant for both $B$ and $R$-band images, we performed aperture photometry to obtain 
semi-major axis radial intensity growth curves. A refined sky determination 
was achieved by slightly varying the background level (at zero) and finding the 
growth curve that converged best to a plateau out to the edge of the image. The 
corresponding deviation from zero was subtracted from the image. The other growth curves 
obtained from sky variations provided an estimate of the sky uncertainty (col.~[5] of Table \ref{tbl4}). The 
total magnitude $m_{\rm T}$ of the galaxy was derived from the asymptotic intensity 
of the growth curve. At half of the maximum we read off the half-light (``effective'') 
radius, $r_{\rm eff}$, and calculated 
the mean surface brightness within $r_{\rm eff}$: the ``effective surface brightness'' 
$\langle \mu\rangle_{\rm eff}$. All photometric and structure parameters for the galaxies 
are listed in Table~\ref{tbl9}. Moreover, we give the values for the galactic extinction 
in each passband (col.~[3]) and the absolute magnitudes of the galaxies (col.~[5]) based on
the distances given in Table~\ref{tbl5}.  

Differentiating the growth curves yielded the $BR$ surface brightness profiles (Fig.\ref{fig6}). 
As characteristic for dE galaxies (Binggeli \& Cameron 1991), we found that their light 
profiles deviate from an exponential (straight line). While UGC4998, the most luminous 
dE in our sample has a cuspy profile in the inner region, all other intrinsically fainter 
dEs exhibit a central decrement relative to an exponential law. This trend is known as the shape 
parameter - luminosity relation (Davies et al.~1988; Jerjen \& Binggeli 1997; JBF00). A 
successful way to quantify the observed variety of surface brightness shapes is 
offered by the S\'ersic profiles (S\'ersic 1968): $\mu(r)=\mu_0+1.086(r/r_0)^n$, 
with a free shape parameter $n$. This family of analytical functions encompasses both, 
the $R^{1/4}$-law and the exponential law ($n=1$). The best-fitting S\'ersic profiles to 
our data are plotted in Fig.\ref{fig6} as solid lines. The corresponding model parameters, 
i.e. the scale length $r_0$, the central surface brightness $\mu_0$, and the shape 
parameter $n$ are listed in Table~\ref{tbl9}. The quoted uncertainties are the profile 
fitting errors.

Fig.\ref{fig7} shows the azimuthally averaged radial $(B-R)_0$ colour gradients for 
our dwarfs out to a corresponding isophotal radius of $\mu_B=28.5$\,mag\,arcsec$^{-2}$. 
The results are consistent with previous findings (e.g.~Patterson \& Thuan 1996) that colour 
gradients of early-type dwarfs are generally small.  

\begin{figure}
\centering
\includegraphics[width=7cm]{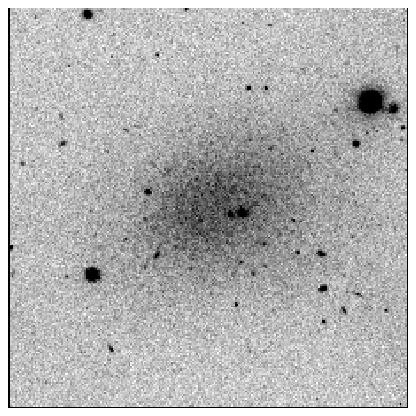}
\includegraphics[width=7cm]{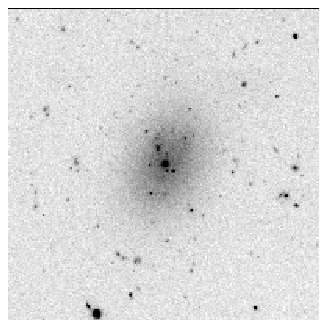}
\caption{$R$-band images of DDO 71 (top panel) and UGC 7356 (bottom panel) 
obtained at the Nordic Optical Telescope with a seeing of $0\farcs8$ and 
$0\farcs9$, respectively. The proposed ``off-centre'' nucleus 11\arcsec~West 
of the centre of DDO 71 is resolved on our image into two stellar-like objects. 
The fainter one is a GC candidate as inferred from its photometric properties. 
Another candidate is the bright, centrally located nucleus of UGC 7356. The 
FOV is $3\farcm0 \times 3\farcm0$. North is up, East to the left. 
}
\label{fig8}
\end{figure}

\begin{figure*}
\centering
\includegraphics[width=17cm]{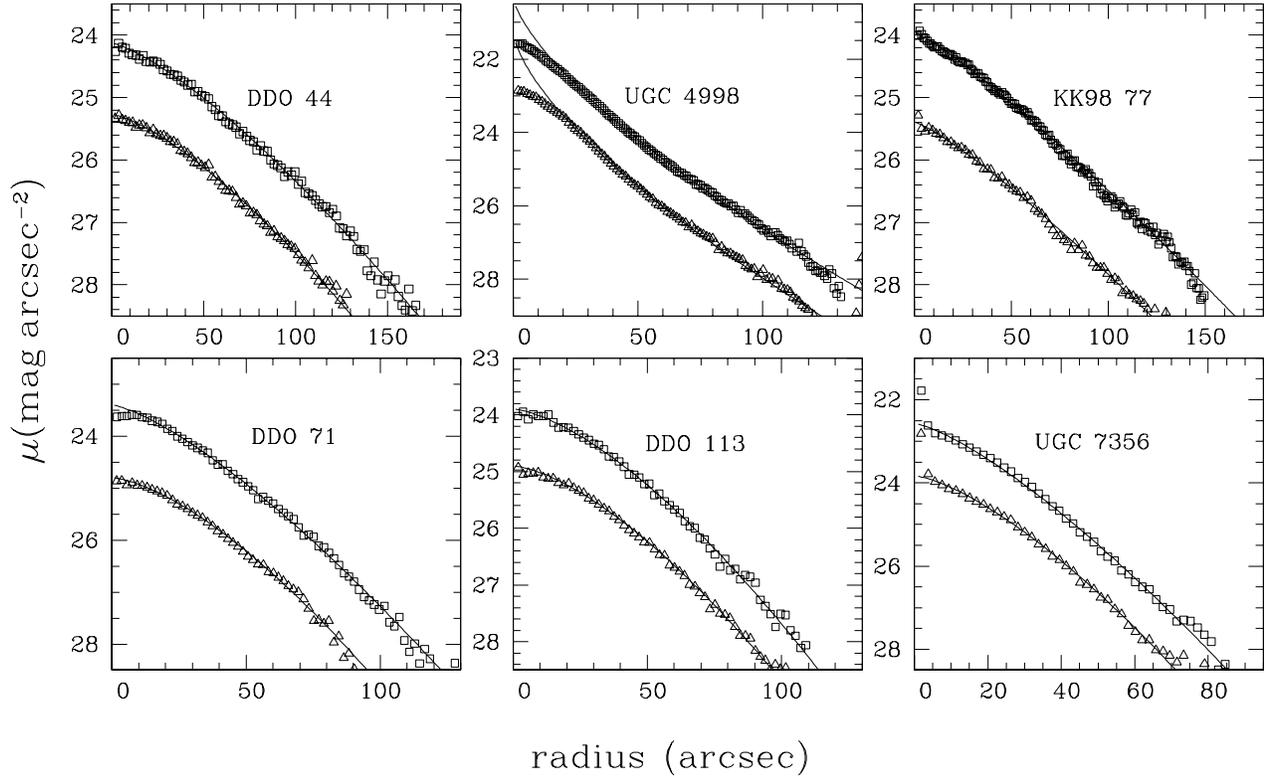}
\caption{Azimuthally averaged surface brightness profiles in the $B$ 
(triangles) and $R$-band (boxes) for the six dwarf ellipticals. The 
best-fitting S\'ersic profiles are plotted as solid lines.
\label{fig6}}
\end{figure*}

\begin{figure*}
\centering
\includegraphics[width=17cm]{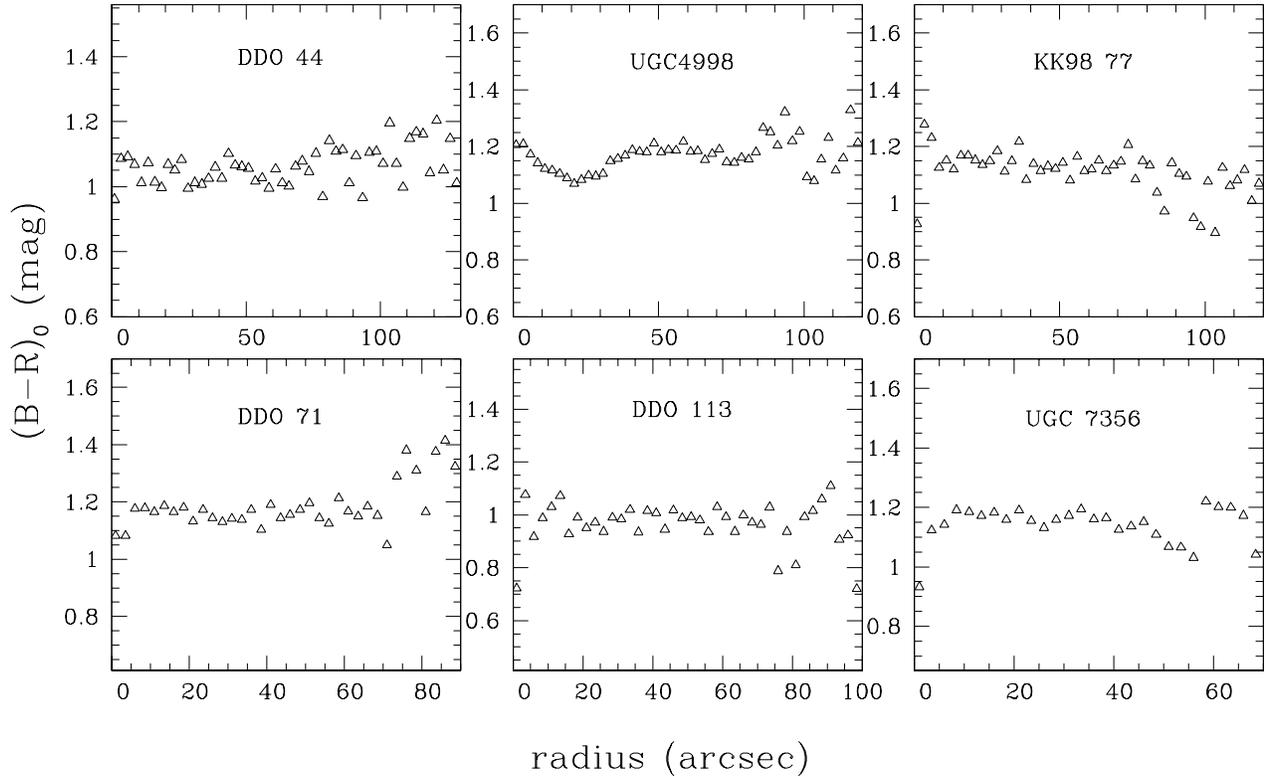}
\caption{Azimuthally averaged $(B-R)_0$ colour profiles of the six dwarf 
ellipticals out to an isophotal radius of $\mu_B=28.5$\,mag\,arcsec$^{-2}$.
\label{fig7}}
\end{figure*}

Finally, we like to draw the attention to a feature in DDO 71: many luminous dE galaxies brighter
than $M_B\approx -14$ have a nucleus, a centrally located object that is possibly 
a massive GC formed in or fallen into the galaxy core region (for a review see Ferguson 
\& Binggeli 1994). Among the nuclei, about 20\% are found to be significantly displaced from 
the galaxy centre as defined by the overall light distribution (Binggeli et al.~2000). DDO 
71 is approximately 1.5\,mag fainter than the mentioned reference magnitude but first images 
suggested that the dwarf may have a nucleus about $11$\,arcsec to the West of the centre 
(Bremnes et al.~1998). Our CCD image resolved this feature into two star-like 
objects (top panel of Fig.\ref{fig8}). K00 suggested that the object to the East is a GC 
candidate (the possible galaxy nucleus) as inferred from the $VI$ colour and magnitudes. 
Table~\ref{tbl8} gives the corresponding basic $BR$ photometry: the total apparent $B$ 
magnitude, $(B-R)$ colour after correction for Galactic reddening, the measured central $B$ 
surface brightness, and the total reddening-corrected absolute $B$ magnitude. Furthermore, we 
present in Table~\ref{tbl8} the photometry for the nucleus of UGC 7356 (bottom panel of 
Fig.\ref{fig8}). This nucleus is closely situated to the galaxy centre and another 
GC candidate. If confirmed as a GC it would have an absolute magnitude of M$^0_{\rm B}=-10.04$ 
and thus would be comparable in luminosity with the brightest Galactic GCs (Harris 1996).

\begin{table}
\caption{Globular cluster candidates \label{tbl8}}
\begin{tabular}{crr} \\ \hline \hline
 Parameter  &DDO 71 & UGC 7356 \\
\hline
R.A.(J2000.0)     &10 05 07.4 & 12 19 09.1 \\             
Decl.(J2000.0)    &66 33 28.7 & 47 05 23.3 \\
   $B_T$          & 21.80  &  19.16   \\    
  $(B-R)_0$       &  1.34  &   1.08   \\ 
$\mu_{0,{B}}$     & 25.87  &  23.55   \\ 
 M$^0_{B}$        &$-6.33$ &  $-$10.04\\ 
\hline \hline
\end{tabular}
\end{table}

\begin{table*}
\small
\caption{Integrated $BR$ photometry and S\'ersic parameters \label{tbl9}}
\begin{tabular}{lccccccccc}
\hline \hline
        &          &  A    & m$_{\rm T}$ & M$^0_{\rm T}$ & r$_{\rm eff}$ & $\langle \mu \rangle_{\rm eff}$ & r$_0$    &   
$\mu_0$              & \\
   Name &  F & (mag)  & (mag) & (mag)       & (arcsec)      &  (mag arcsec$^{-2}$)           & (arcsec) &        
(mag arcsec$^{-2}$) & n  \\
(1) & (2)   & (3)  & (4)       & (5)      & (6)    & (7)    & (8)      & (9) & (10) \\
\hline
DDO 44   & B      &  0.18$\pm0.03$ & 15.44$\pm0.05$ & $-12.26$ & 51.7$\pm1.6$ & 26.00$\pm0.03$ & 59.7$\pm2.6$ & 25.27$\pm0.05$ & 1.39$\pm0.07$ \\
DDO 44   & R      &  0.11$\pm0.02$ & 14.31$\pm0.03$ & $-13.32$ & 51.8$\pm0.7$ & 24.87$\pm0.02$ & 58.7$\pm1.6$ & 24.14$\pm0.03$ & 1.33$\pm0.04$ \\
UGC 4998 & B      &  0.24$\pm0.04$ & 14.68$\pm0.05$ & $-15.67$ & 24.8$\pm0.4$ & 23.65$\pm0.03$ &  6.6$\pm3.1$ & 21.25$\pm0.66$ & 0.67$\pm0.09$ \\
UGC 4998 & R      &  0.16$\pm0.03$ & 13.45$\pm0.03$ & $-16.82$ & 26.2$\pm0.2$ & 22.54$\pm0.03$ &  8.1$\pm2.0$ & 20.30$\pm0.13$ & 0.70$\pm0.05$ \\
KK98 77  & B      &  0.62$\pm0.10$ & 15.54$\pm0.05$ & $-12.80$ & 54.4$\pm1.5$ & 26.21$\pm0.03$ & 50.9$\pm1.9$ & 25.39$\pm0.05$ & 1.20$\pm0.04$ \\
KK98 77  & R      &  0.39$\pm0.06$ & 14.20$\pm0.04$ & $-13.90$ & 52.4$\pm0.9$ & 24.79$\pm0.03$ & 48.1$\pm3.3$ & 23.95$\pm0.07$ & 1.16$\pm0.07$ \\ 
DDO 71   & B      &  0.41$\pm0.07$ & 15.68$\pm0.05$ & $-12.45$ & 35.0$\pm0.9$ & 25.40$\pm0.03$ & 42.0$\pm1.4$ & 24.81$\pm0.04$ & 1.50$\pm0.07$ \\ 
DDO 71   & R      &  0.26$\pm0.04$ & 14.32$\pm0.04$ & $-13.66$ & 36.2$\pm0.8$ & 24.11$\pm0.03$ & 39.0$\pm1.8$ & 23.40$\pm0.07$ & 1.34$\pm0.06$ \\
DDO 113  & B      &  0.09$\pm0.02$ & 15.86$\pm0.04$ & $-11.67$ & 34.3$\pm0.8$ & 25.53$\pm0.03$ & 43.7$\pm2.7$ & 24.92$\pm0.08$ & 1.51$\pm0.13$ \\
DDO 113  & R      &  0.06$\pm0.01$ & 14.79$\pm0.05$ & $-12.71$ & 35.2$\pm1.1$ & 24.52$\pm0.03$ & 43.51$\pm3.6$& 23.90$\pm0.13$ & 1.50$\pm0.12$ \\
UGC 7356 & B      &  0.08$\pm0.01$ & 15.67$\pm0.04$ & $-13.53$ & 22.3$\pm0.5$ & 24.40$\pm0.03$ & 25.71$\pm1.3$& 23.83$\pm0.06$ & 1.43$\pm0.09$ \\  
UGC 7356 & R      &  0.05$\pm0.01$ & 14.49$\pm0.04$ & $-14.68$ & 22.0$\pm0.4$ & 23.20$\pm0.03$ & 26.53$\pm2.4$& 22.68$\pm0.14$ & 1.51$\pm0.14$ \\ 
\hline
\hline
\end{tabular}
\end{table*}

\section{Summary and Conclusions}
We have analysed $BR$-band CCD images of six nearby dwarf ellipticals 
and measured stellar $R$-band surface brightness fluctuation magnitudes 
$\overline{m}_R$ and $(B-R)$ colours in 29 galaxy field. We 
combined our photometry data with independent TRGB distances for the 
three dwarfs DDO 44, KK98 77, and DDO 71, and compared the resulting 
empirical $(B-R)_0-\overline{M}_R$ relation with the theoretical predictions 
for old, metal poor stellar populations based on Worthey's population 
synthesis models and the evolutionary tracks from the Padova library.
While the general colour dependency of $\overline{M}_R$ could be confirmed, we
found a systematic shift of $0.13\pm(0.02)$\,mag between observed 
and model fluctuation magnitudes in the studied colour range 
$1.0<(B-R)_0< 1.25$. Once a simple offset correction was applied to match 
the empirical SBF zero point, the parabolic branch of the theoretical 
$(B-R)-\overline{M}_R$ relation given by the form $\overline{M}_R = 
1.89\,[(B-R)_0-0.77]^2-1.39$ followed closely the empirical results 
and thus provided the first semiempirical calibration of the stellar 
fluctuation magnitudes in dEs as a function of the distance independent 
$(B-R)$ colour. The results indicated that the improved calibration, resting on the 
tip magnitude of the RGB stars and Worthey+Padova, can be used as 
a distance indicator for genuine dwarf elliptical galaxies with an 
estimated accuracy of $\approx 10$\%. Two points are essential 
for the SBF method to work successfully. Firstly, the stellar fluctuation
magnitude has to be measured in a number of fields in a galaxy. 
Colour differences between fields and the resulting $\overline{m}_R$ variation 
can be used to determine on which calibration branch a galaxy lies. Secondly, the 
application of the SBF method must focus on pure breed dwarf ellipticals. 
Intermediate type (dE/Irr) dwarfs show evidence of more recent star-formation 
activities and have more complex star formation histories than genuine dEs. 
This prevent a reliable interpretation of the fluctuation magnitudes as 
demonstrated in a previous study (Jerjen \& Rejkuba 2000),   

Taking advantage of the new calibration of the SBF method for dEs, we 
derived first distances for the three early-type dwarfs UGC 4998, DDO 113 
and UGC 7356. UGC 4998 could be confirmed as a stellar system in the 
background of the M 81 group situated right at the periphery of the 
10\,Mpc sphere at 10.5\,Mpc. DDO 113 and UGC 7356, are both known 
members of the spatially extended Canes Venatici I cloud based on 
their redshifts. We found them at 3.1\,Mpc and 6.7\,Mpc, respectively. 

The example of UGC 4998 demonstrated the great potential of the SBF method 
to obtain accurate distances for dwarf elliptical galaxies as far away 
as 10\,Mpc. The modest requirement of two hours of imaging under good seeing 
conditions at a 2.5m-class ground-based telescope opens up the possibility to 
measure distances to all known early-type dwarfs in the vicinity of the Local 
Group and to newly discovered dE candidates in an efficient and simple way. 

Several steps could be undertaken to further improve the method. SBF data for 
more calibrator galaxies are needed to explore the empirical $(B-R)_0-\overline{M}_R$ 
relation over a larger range in colour. Particularly useful will be galaxies that 
populate the linear branch of the relation. They can help to pin down the 
empirical zero point and slope of the second theoretical branch. In a next step, 
the $(B-R)_0-\overline{M}_R$ diagram may be tested as a tool to estimate 
the spread in age and metallicity over a dwarf galaxy's surface.

In the last part of our study, we presented $BR$ surface brightness 
and $(B-R)$ colour profiles for the dwarfs. Radial colour gradients 
were found to be generally small and the galaxy light profiles follow the shape 
parameter -- luminosity relation for dEs as previously observed in 
other dE samples (JFB00). Two GC candidates in DDO 71 and UGC 7356 
have been discussed and photometric data provided.  
 
\begin{acknowledgements}
It is a pleasure to thank B.~Binggeli and G.~Da Costa for useful discussions.
We are grateful to the referee Dr.~R.I.~Thompson whose comments helped to improve 
the paper. The Nordic Optical Telescope is operated on the island of La Palma jointly by 
Denmark, Finland, Iceland, Norway, and Sweden, in the Spanish Observatorio del 
Roque de los Muchachos of the Instituto de Astrofisica de Canarias.
\end{acknowledgements}

\end{document}